\documentclass[preprint,aps]{revtex4}
\usepackage{graphicx}
\usepackage{dcolumn}
\usepackage{bm}

\def\be{\begin{equation}}
\def\ee{\end{equation}}
\def\beq{\begin{eqnarray}}
\def\eeq{\end{eqnarray}}

\def\bay{\begin{array}}
\def\eay{\end{array}}

\def\test{{\raisebox{-1.0ex}{\scriptsize$r\rightarrow 0$} \atop \raisebox{0.8ex}{\scriptsize$t
\rightarrow 0$}}}
\begin{document}

\preprint{CIRI/01-swkg02}
\title{Naked Singularities in Spherically Symmetric, Self-Similar Spacetimes}

\author{Sanjay M. Wagh}
\affiliation{Central India Research
Institute, Post Box 606, Laxminagar, Nagpur 440 022, India\\
E-mail:ciri@vsnl.com}

\author{Keshlan S. Govinder}
\affiliation{School of Mathematical and Statistical
Sciences, University of Natal, Durban 4041, South Africa \\
E-mail: govinder@nu.ac.za\\ }
\bigskip

\date{December 22, 2001}

\begin{abstract}
We show that all known naked singularities in spherically
symmetric self-similar spacetimes arise as a result of singular
initial matter distribution.  This is a result of the peculiarity
of the coordinate transformation that takes these spacetimes into
a separable form. Therefore, these examples of naked singularities
are of no apparent consequence to astrophysical observations or
theories.
 \\

\centerline{Submitted to: Physical Review Letters}
\end{abstract}

\pacs{04.20.-q, 04.20.Cv, 95.30.5f}%
\maketitle

\newpage
In recent years, many workers appear to consider naked or visible
spacetime singularities to be of serious astrophysical
significance \cite{djm}.  This is in  contrast to the general
research atmosphere of most of the second half of the twentieth
century where the emphasis was on proving that naked singularities
are ``unphysical'' in the sense and spirit of the Cosmic
Censorship Hypothesis (CCH). This ``paradigm shift'' can be
attributed to different, specific, known examples of naked
singularities.

It is noteworthy that such solutions have been found to the highly
nonlinear coupled partial differential equations as are the field
equations of General Relativity for some realistic equations of
state for the matter in the spacetime. There is however a general
consensus that these examples,  being mostly spherically
symmetrical in nature, do not counter the spirit of the Cosmic
Censorship Hypothesis (CCH). It is then hoped that these examples
of naked singularities, apart from their special symmetry, may
help sharpen the statement of the CCH that is as yet unproven.

Many of the known examples occur in the highly specialized
spherically symmetric, self-similar spacetimes (SSSSS). Such
spacetimes admit a homothetic Killing vector. All points along the
integral curves of the homothetic Killing vector field are
equivalent. Consequently, for a self-similar spacetime the field
equations of General Relativity reduce to ordinary differential
equations.

It is the purpose of this letter, to re-examine the applicability
of the CCH as regards SSSSS notwithstanding the results in, for
example, \cite{pirangrf, nolan}.

The requirement that the general spherically symmetric line
element \begin{widetext} \be ds^2 \,=\, - e^{2\nu(t,r)} dt^2 \,+\,
e^{2\lambda(t,r)} dr^2 \,+\, Y^{2}(t,r)(d\theta^2 + \sin^2 \theta
d\phi^2) \label{met1} \ee \end{widetext} admits a homothetic
Killing vector of the form \be X^a\;=\;(0,f(r,t),0,0) \label{vec1}
\ee reduces  \cite{prl1} this metric {\em uniquely\/} to:
\begin{widetext} \be
ds^2\;=\;-\,y^2(r)\,dt^2\;+\;\gamma^2\,B^2(t)\,\left(y'\right)
^2\,dr^2\;+\;y^2(r)\,Y^2(t) \left[ \,
d\theta^2\;+\;\sin^2{\theta}\,d\phi^2 \,\right] \label{met}\ee
\end{widetext} with an overhead prime denoting differentiation with
respect to $r$, $f(r,t)$ in (\ref{vec1}) now becoming only a
function of $r$, viz. \be f(r,t)=\frac{y}{\gamma y'}\label{fofr}
\ee where $\gamma$ is a constant and we have absorbed the temporal
coefficient of $dt^2$ by a redefinition of $t$. This spacetime
admits an energy density of the form \be \rho =
\frac{1}{y^2}\left(\frac{\dot{Y}^2}{Y^2} + 2 \frac{\dot{B}}{B}
\frac{\dot{Y}}{Y} + \frac{1}{Y^2} - \frac{1}{\gamma^2 B^2}\right)
\label{sepdens}  \ee Clearly, the density is non-singular/regular
for $y(r)\neq 0$ for all $r$. We note that the density is a
decreasing function of $r$ when $y'>0$ and an increasing function
of $r$ when $y'<0$. (We also note that for constant $y$, the
spacetime is homogeneous and for $y=\infty$, the spacetime is a
vacuum solution.) The singularity develops in the collapse of this
regular density distribution of matter as a result of its temporal
evolution, that is, when the temporal part in (\ref{sepdens})
becomes infinite at some $t=t_0$. By symmetry considerations, this
singularity must, of necessity, lie at $r=0$ when $t=t_0$.

As can be shown \cite{collapse}, (\ref{met}) does {\it not} admit
any naked singularities, provided the initial data is regular.  We
recall here some of the relevant results from \cite{collapse} for
the sake of completeness. From (\ref{met}) it follows that the
radial null geodesic satisfies \be \frac{dt}{dr}=\pm\, \gamma
\frac{y'}{y} B \label{rng} \ee

A non-singular density profile requires $y(r) \neq 0$ for all $r$.
In this case, we shall have
\[ \lim_{r \rightarrow 0}\;\frac{y^{\prime}}{y} = \ell_o \]
say, where $\ell_o$ is the corresponding finite limiting value.
(Note that $\ell_0$ can be positive or negative depending on the
sign of $y'$.)

Now, for the geodesic tangent to be uniquely defined and to exist
at the singular point, $t=0$ and $r=0$, the following must hold so
that an outgoing, future-directed photon trajectory exists at the
singularity: \be \lim_{\test}\;\; \frac{t}{r} = \lim_{\test}\;\;
\frac{dt}{dr} = X_o \label{limits} \ee where $X_o$ is required to
be {\em real\/} and {\em positive}. As we approach the singular
point, we have \be X_o=\pm\gamma\,\,\lim_{\test}
\,\,\frac{y^{\prime}}{y} \,B(t) \;=\;\pm\gamma\,\,\lim_{ t
\rightarrow 0} \;\ell_o \,B(t) \label{ourlimit} \ee Clearly, the
limit in (\ref{ourlimit}) implies that $X_o =0$. Therefore, in the
limit of the singularity of (\ref{met}), that is,
$t\rightarrow0,r\rightarrow0$ there does not exist a radially
outgoing tangent to the radial null geodesic. Hence, the spacetime
singularity is not naked or visible if the initial density profile
corresponds to a density distribution that is non-singular.

The obvious question is the reconciliation of this result with the
proliferation of naked singularities in SSSSS widely reported in
the literature.

In this connection, we note that the usual form of the homothetic
Killing vector is taken to be \cite{psj}: \be
\bar{X}^a\;=\;(T,R,0,0) \label{vec2} \ee Indeed most authors
impose (\ref{vec2}) on (\ref{met1}) before performing any further
analysis.  As a result, the metric they focus on is
\begin{widetext} \be
ds^2\;=\;-\,P\left(\frac{T}{R}\right)^2\,dT^2\;+\;Q\left(\frac{T}
{R}\right)^2\,dR^2\;+\; R^2\,S\left(\frac{T}{R}\right)^2\,
\left[d\theta^2\;+\;\sin^2{\theta}\,d\phi^2 \,\right]
\label{ssmet} \ee \end{widetext} where $P, Q, S$ are the metric
functions of the self-similarity variable $T/R$. In the discussion
of self-similar collapse as considered by most of these authors,
one is therefore led to consider the singularity of (\ref{ssmet})
at $R=0, T=0$.

We emphasize that any vector of the form (\ref{vec1}) can be
transformed into (\ref{vec2}) via the coordinate transformation
\begin{widetext}
\be R = l(t) \exp\left(\int f^{-1} d r\right) \qquad T = k(t)
\exp\left(\int f^{-1} d r\right) \label{sstrans}\ee
\end{widetext}
which, in view of (\ref{fofr}), reduces to \be R = l(t) y^{\gamma}(r)
\qquad T = k(t) y^{\gamma}(r) \label{finaltrans} \ee ({\bf Note:}
If we invoke (\ref{sstrans}), the resulting metric will not be
diagonal. The imposition of diagonality of the metric will require
a relationship between $l(t)$ and $k(t)$. Such a relation can
always be imposed.)

As we noted above, a relationship exists between $l(t)$ and
$k(t)$. Thus $l(t)=0$ for (\ref{met}) corresponds to both $R=0$
{\it and} $T=0$ for (\ref{ssmet}). There is thus no constraint on
the radial function $y(r)$ in (\ref{met}). This will correspond to
those sectors in SSSSS in which no naked singularities arise.

However, we also note that $R=0$ and $T=0$ also corresponds to
$y(r)=0$. We have seen \cite{collapse} that the only case of
obtaining a naked singularity for (\ref{met}) is that of assuming
$y(r)=0$ for $r=0$ which makes (\ref{sepdens}) initially singular.
This corresponds to those sectors of SSSSS in which naked
singularities do arise. It is therefore not surprising that a
naked singularity arises in (\ref{ssmet}) as one assumes a naked
singularity at the outset. This is also evident from the density
distribution for (\ref{ssmet}) \be \rho = \frac{1}{R^2}
\eta\left(\frac{T}{R}\right) \label{ssdens} \ee also being
initially singular for $R=0$. We emphasize that this is true for
{\em any\/} equation of state for matter in the spacetime.

In conclusion, we note that the metric (\ref{met}) contains all
spacetimes admitting a homothetic Killing vector, that have
vanishing or non-vanishing shear and/or energy flux provided the
transformation (\ref{finaltrans}) is non-singular, ie when
$y'\neq0$ (inhomogeneous) and $y'\neq \infty$ (non-vacuum) for all
$r$. We therefore conclude that naked singularities do not arise
in SSSSS for regular initial data.  The known examples of naked
singularities in SSSSS must correspond to singular initial data
for matter fields. Consequently, these examples of naked
singularities neither form counter-examples to CCH nor provide any
implications vis-a-vis astrophysical observations. In fact, the
metric (\ref{met}) not possessing any naked singularities for
regular initial data for matter fields is in complete agreement
with the strong CCH \cite{penrose}.

\section*{Acknowledgements}

KSG thanks the University of Natal and the National Research
Foundation for ongoing support.  He also thanks CIRI for their
kind hospitality during the course of this work.

\end{document}